\documentclass[webpdf,contemporary,large,namedate]{oup-authoring-template}%

\DeclareMathOperator*{\argmax}{argmax}

\usepackage{algorithmicx}
\usepackage{lmodern}
\usepackage{setspace}

\begin{document}
\journaltitle{TBD}
\DOI{TBD}
\copyrightyear{2025}
\pubyear{2025}
\access{Advance Access Publication Date: Day Month Year}
\appnotes{Preprint}
\firstpage{1}

\title[Finding low-complexity sequences]{Finding low-complexity DNA sequences with longdust}
\author[1,2,3,$\ast$]{Heng Li\ORCID{0000-0003-4874-2874}}
\author[4]{Brian Li}
\address[1]{Department of Biomedical Informatics, Harvard Medical School, 10 Shattuck St, Boston, MA 02215, USA}
\address[2]{Department of Data Science, Dana-Farber Cancer Institute, 450 Brookline Ave, Boston, MA 02215, USA}
\address[3]{Broad Insitute of MIT and Harvard, 415 Main St, Cambridge, MA 02142, USA}
\address[4]{Commonwealth School, Boston, MA 02116, USA}
\corresp[$\ast$]{Corresponding author. \href{mailto:hli@ds.dfci.harvard.edu}{hli@ds.dfci.harvard.edu}}


\abstract{
\sffamily\footnotesize
\textbf{Motivation:}
Low-complexity (LC) DNA sequences are compositionally repetitive sequences
that are often associated with spurious homologous matches and variant calling artifacts.
While algorithms for identifying LC sequences exist,
they either do not define complexity mathematically or are inefficient with long or variable context windows.
\vspace{0.5em}\\
\textbf{Results:}
Longdust is a new algorithm that efficiently identifies long LC sequences including centromeric satellite
and tandem repeats with moderately long motifs.
It defines string complexity by statistically modeling the $k$-mer count distribution
with the parameters: the $k$-mer length, the context window size and a threshold on complexity.
Longdust exhibits high performance on real data and high consistency with existing methods.
\vspace{0.5em}\\
\textbf{Availability and implementation:}
\url{https://github.com/lh3/longdust}
}

\maketitle

\section{Introduction}

A string is of low complexity (LC) if it is repetitive in composition.
LC strings may lead to spurious homologous matches~\citep{Morgulis:2006aa} and cause variant calling artifacts~\citep{Li:2014aa,Qin:2025aa}.
In computer science, string complexity can be measured with Kolmogorov complexity~\citep{DBLP:journals/tcs/Kolmogorov98,Silva:2022aa},
Lempel-Ziv complexity~\citep{DBLP:journals/tit/LempelZ76,Orlov:2004aa}
or Shannon entropy~\citep{DBLP:journals/bstj/Shannon48}.
In genomics, topological entropy~\citep{Crochemore:1999aa},
linguistic complexity~\citep{Troyanskaya:2002aa}
and Markov additive process~\citep{10.2307/27595858} have also been explored.
A complication in finding LC regions is
that we want to identify boundaries of local LC regions,
instead of classifying an entire string as LC or not.
Measuring complexity with fixed-sized sliding windows is not adequate, either,
as we could not pinpoint the boundaries with such methods.

SDUST~\citep{Morgulis:2006aa} is a popular algorithm to identify local LC regions.
It defines string complexity by running 3-mer counts and finds exact solutions under this definition.
However, with the $O(w^3L)$ time complexity, where $w$ is the window size and $L$ is the genome length,
SDUST is inefficient given a large $w$.
In addition, the complexity scoring function used by SDUST grows quadratically with the string length.
It tends to classify long strings as LC over short ones.
A large window size $w$ would make this worse.
As a result, SDUST is not suitable for finding satellite or tandem repeats with long motifs.

In genomes, LC strings are often tandemly repetitive due to DNA polymerase slippage.
Such tandem repeats can be identified with tandem repeat finding algorithms such as
TRF~\citep{Benson:1999aa}, ULTRA~\citep{Olson:2024aa} and pytrf~\citep{Du:2025aa}.
They report repeat motifs and copy numbers, which are useful in downstream analyses.
TANTAN~\citep{Frith:2011aa} models tandem patterns but can often identify general LC regions in practice.
Unlike SDUST, these algorithms do not mathematically define string complexity.

Inspired by SDUST, we sought an alternative way to define the $k$-mer complexity of a string
and to identify local LC regions in a genome.
Our complexity scoring function is based on a statistical model of $k$-mer count distribution
and our algorithm is practically close to $O(wL)$ in time complexity,
enabling the efficient identification of LC strings in long context windows.

\section{Methods}

Similar to SDUST~\citep{Morgulis:2006aa}, we define the complexity of a DNA string with a function of $k$-mer counts.
In this section, we will model the $k$-mer count distribution of random strings with each $k$-mer occurring at equal frequency
and describe the complexity scoring function and the condition on bounding LC substrings in a long string.
We will then generalize the model to unequal $k$-mer frequency.
We will also discuss alternative scoring functions such as the SDUST scoring and Shannon entropy.

\subsection{Notations}

Let $\Sigma=\{{\tt A},{\tt C},{\tt G},{\tt T}\}$ be the DNA alphabet,
$x\in\Sigma^*$ is a DNA string and $|x|$ is its length.
$t\in\Sigma^k$ is a $k$-mer.
For $|x|\ge k$, $c_x(t)$ is the occurrence of $k$-mer $t$ in $x$;
$\ell(x)=\sum_t c_x(t)=|x|-k+1$ is the total number of $k$-mers in $x$.
$\vec{c}_x$, of size $4^k$, denotes the count array over all $k$-mers.

In this article, we assume there is one long genome string of length $L$.
We use closed interval $[i,j]$ to denote the substring starting at $i$
and ending at $j$, including the end points $i$ and $j$.
We may use ``interval'' and ``substring'' interchangeably.

\subsection{Modeling $k$-mer counts}\label{sec:model}

For a $k$-mer $t\in\Sigma^k$, $q_t$ is its frequency with $\sum_t q_t=1$.
$\vec{c}_x$ follows a multinomial distribution:
$$
P(\vec{c}_x;\vec{q})=\ell(x)!\prod_{t\in\Sigma^k}\frac{q_t^{c_x(t)}}{c_x(t)!}
$$
For convenience, introduce $r_t\triangleq 4^kq_t$.
We have
\begin{equation}\label{eq:P}
\log P(\vec{c}_x;\vec{q})=g(\ell(x))-\sum_t\log c_x(t)!+\sum_tc_x(t)\log r_t
\end{equation}
where
$$
g(\ell)=\log \ell!-k\ell\log4
$$
is a function of string length but does not depend on $k$-mer counts.

To get an intuition about the effect of low-complexity strings,
suppose $q_t=1/4^k$ and $\ell(x)\ll4^k$.
In this case, the last term in Eq.~(\ref{eq:P}) is 0
and $c_x(t)$ is mostly 0 or 1 for a random string.
$\log P(\vec{c}_x)$ will be close $g(\ell(x))$.
Given an LC string of the same length,
we will see more $c_x(t)$ of 2 or higher, which will reduce $\log P(\vec{c}_x)$.
Thus the probability of an LC string is lower under this model.

Although $\log P(\vec{c}_x)$ can be used to compare the complexity of strings of the same length,
it does not work well for strings of different lengths because $\log P(\vec{c}_x)$ decreases with $\ell(x)$.
We would like to scale it to $Q(\vec{c}_x)$ such that
$Q$ approaches 0 given a random string.
We note that the mean of $\log P(\vec{c}_x)$
$$
H(\ell)=\sum_{\vec{c}_x}P(\vec{c}_x)\log P(\vec{c}_x)
$$
is the negative entropy of $P(\vec{c}_x)$, which is
$$
H(\ell)=\log\ell!+\ell\sum_t q_t\log q_t-f(\ell)=g(\ell)+\ell\sum_t q_t\log r_t-f(\ell)
$$
where
$$
f(\ell;\vec{q})\triangleq\sum_t\sum_{n=0}^{\ell}\binom{\ell}{n}q_t^n(1-q_t)^{\ell-n}\log n!
$$
Under the Poisson approximation,
\begin{equation}\label{eq:f}
f(\ell;\vec{q})\approx\sum_t e^{-\ell q_t}\sum_{n=0}^{\infty}\log n!\cdot\frac{(\ell q_t)^n}{n!}
\end{equation}
Now introduce
\begin{eqnarray*}
\tilde{Q}(\vec{c}_x)&\triangleq&H(\ell)-\log P(\vec{c}_x)\\
&=&\sum_t\log c_x(t)!-f(\ell(x))+\sum_t\big[\ell(x) q_t-c_x(t)\big]\log r_t
\end{eqnarray*}
$\tilde{Q}(\vec{c}_x)$ approaches 0 given a random string $x$ regardless of its length.
We can thus compare strings of different lengths.
However, $\tilde{Q}$ measures both $k$-mer repetitiveness in the first term
and $k$-mer usage in the last term.
It is possible for $\tilde{Q}$ to reach a large value if string $x$ is composed of rare unique $k$-mers.
To cancel the effect of rare $k$-mers, we ignore the last term and use
$$
Q(\vec{c}_x;\vec{q})\triangleq\sum_t\log c_x(t)!-f(\ell(x);\vec{q})
$$
to measure the complexity of string $x$.
$Q(\vec{c}_x)$ also approaches 0 given a random string $x$
because $\tilde{Q}(\vec{c}_x)$ approaches 0 and $c_x(t)$ approaches $\ell(x)q_t$.
$Q(\vec{c}_x)$ is higher for LC strings.

\begin{figure}[tb]
\centering
\includegraphics[width=\columnwidth]{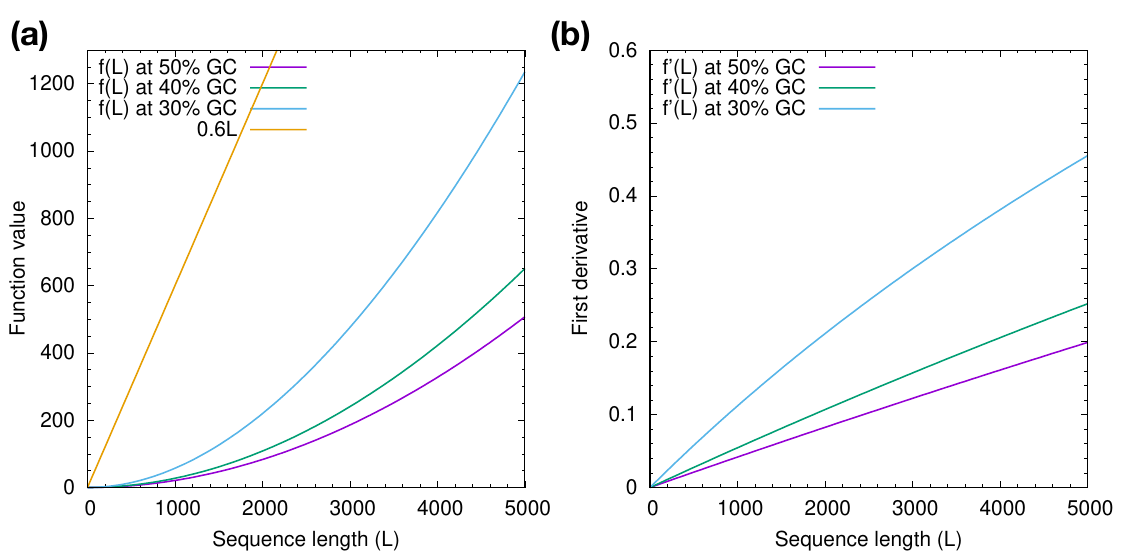}
\caption{Scaling function.
{\bf (a)} Function value of $f(\ell)$ in Eq.(\ref{eq:f}) at $k=7$.
{\bf (b)} First derivative. Equivalent to per-base contribution.}\label{fig:f}
\end{figure}

In practice, $k$-mer frequency $\vec{q}$ is computed from genome-wide GC content.
Low or high GC content leads to elevated $f(\ell)$ (Fig.~\ref{fig:f}),
which is expected as $c_x(t)$ grows faster with more biased GC content.

\subsection{Scoring low-complexity intervals}

To put a threshold on the complexity, we finally use the following function to score string complexity:
\begin{equation}\label{eq:S}
S(\vec{c}_x)\triangleq Q(\vec{c}_x)-T\cdot\ell(x)=\sum_t\log c_x(t)!-T\cdot\ell(x)-f(\ell(x))
\end{equation}
Threshold $T$ controls the level of complexity in the output.
It defaults to $0.6$, less than $\log 2$.
At $k=7$, $f(\ell)$ is close to 0 for small $\ell$.
$0.6\ell$ is the larger term within the default window size of 5kb (Fig.~\ref{fig:f}).

If $S(\vec{c}_x)>0$, $x$ is considered to contain an LC substring.
However, we often do not want to classify the entire $x$ as an LC substring in this case
because the concatenation of a highly repetitive sequence and a random sequence
could still lead to a positive score.

Recall that we may use close intervals to represent substrings.
For convenience, we write $S(\vec{c}_{[i,j]})$ as $S(i,j)$.
In implementation, we precalculate $f(\ell)$ and introduce
\begin{eqnarray*}
U(i,j)&\triangleq&\sum_t\log c_{[i,j]}(t)!-T\cdot\ell([i,j])\\
&=&U(i,j-1)+\log c_{[i,j]}([j-k+1,j])-T
\end{eqnarray*}
We can thus compute the complexity scores of all prefixes of $[i,j]$
by scanning each base in the interval from left to right;
we can similarly compute all suffix scores from right to left.

\subsection{Finding low-complexity regions}

We say $x$ is a \emph{perfect LC string} (or \emph{perfect LC interval})
if $S(\vec{c}_x)>0$ and no substring of $x$ is scored higher than $S(\vec{c}_x)$;
say $x$ is a \emph{good LC string} (or \emph{good LC interval})
if $S(\vec{c}_x)>0$ and no prefix or suffix of $x$ is scored higher than $S(\vec{c}_x)$.
We can use $U(i,j)$ above to test if $[i,j]$ is a good LC interval in linear time.
If we apply this method to all intervals up to $w$ in length (5000bp by default),
we can find LC regions of context length up to $w$ in $O(w^2L)$ time.
The union of all good LC intervals marks the LC regions in a genome.

\begin{algorithm}[bt]
	\caption{Find LC interval ending at $j$}\label{algo:LC1}
	\begin{algorithmic}[1]
		\Procedure{FindStart}{$k,w,T,j,c'$}
			\State $B\gets${\sc Backward}$(k,w,T,j,c')$
			\State $j'_{\max}\gets-1$
			\For{$(i,v')\in B$ in the ascending order of $i$}
				\State {\bf continue if} $i<j'_{\max}$\Comment{this is an approximation}
				\State $j'\gets${\sc Forward}$(k,T,i,j,v')$
				\State \Return $i$ {\bf if} $j'=j$\Comment{$[i,j]$ is a good LC interval}
				\State $j'_{\max}\gets\max(j'_{\max},j')$
			\EndFor
			\State \Return $-1$\Comment{No good LC interval ending at $j$}
		\EndProcedure
		\Procedure{Backward}{$k,w,T,j,c'$}
			\State $u\gets 0$; $v_0\gets-1$; $u'\gets 0$
			\State $v_{\max}\gets0$; $i_{\max}\gets-1$; $c\gets[0]$
			\State $B\gets\emptyset$
			\For{$i\gets j$ {\bf to} $\max(j-w+1,k-1)$}\Comment{$i$ is descending}
				\State $t\gets[i-k+1,i]$\Comment{the $k$-mer ending at $i$}
				\State $c[t]\gets c[t]+1$
				\State $u\gets u+\log(c[t])-T$
				\State $v\gets u-f(j-i+1)$\Comment{$v=S(i-k+1,j)$}
				\If{$v<v_0$ {\bf and} $v_0=v_{\max}$}
					\State $B\gets B\cup\{(i+1,v_{\max})\}$\Comment{a candidate start pos}
				\ElsIf{$v\ge v_{\max}$}
					\State $v_{\max}\gets v$; $i_{\max}\gets i$
				\ElsIf{$i_{\max}<0$}
					\State $u'\gets u'+\log(c'[t])-T$\Comment{$c'[t]\triangleq c_{[j-w+1,j]}(t)$}
					\State {\bf break if} $u'<0$\Comment{{\sc Forward}() wouldn't reach $j$}
				\EndIf
				\State $v_0\gets v$
			\EndFor
			\State $B\gets B\cup\{(i_{\max},v_{\max})\}$ {\bf if} $i_{\max}\ge0$
			\State\Return $B$
		\EndProcedure
		\Procedure{Forward}{$k,T,i_0,j,v'_{\max}$}
			\State $u\gets 0$; $v_{\max}\gets0$; $i_{\max}\gets-1$; $c\gets[0]$
			\For{$i\gets i_0$ {\bf to} $j$}
				\State $t\gets[i-k+1,i]$
				\State $c[t]\gets c[t]+1$
				\State $u\gets u+\log(c[t])-T$
				\State $v\gets u-f(i-i_0+1)$
				\If{$v\ge v_{\max}$}
					\State $v_{\max}\gets v$; $i_{\max}\gets i$
				\EndIf
				\State {\bf break if} $v>v'_{\max}$
			\EndFor
			\State\Return $i_{\max}$
		\EndProcedure
	\end{algorithmic}
\end{algorithm}

Algorithm~\ref{algo:LC1} shows a faster way to find a good LC interval ending at $j$.
Function {\sc Backward}() scans backwardly from $j$ to $j-w+1$ to collect candidate start positions (line 22).
Variable $v$ is the complexity score of suffix $[i-k+1,j]$ (line 20).
By the definition of good LC interval, $i$ can only be a candidate start if $v$ is no less than all the suffixes visited before (line 21).
We also ignore a candidate start $i$ if $S(i,j)<S(i-1,j)$
because if $[i-1,j]$ is not a good LC interval, there must exist $i'>i$ such that $S(i',j)>S(i-1,j)<S(i,j)$, so $[i,j]$ would not be a good LC interval, either.
In addition, if suffix $[i,j]$ is enriched with $k$-mers unique in the full window $[j-w+1,j]$,
$[i,j]$ will not be a good LC interval (line 27) as there will exist $j'<j$ such that $S(i,j')>S(i,j)$.
The time complexity of {\sc Backward}() is $O(w)$.

Given a candidate start position $i$,
function {\sc Forward}() returns $j'=\argmax_{i<j'\le j} S(i,j')$.
$[i,j]$ will be a good LC interval if and only if $j'=j$ (Line 7).
We call {\sc Forward}() in the ascending order of candidate start positions (line 4).
We may skip a start position if it is contained in an interval found from previous {\sc Forward}() calls (line 5).
This is an approximation as it is possible for a good LC interval to start in another good interval.
An alternative heuristic is to only apply {\sc Forward}() to the smallest candidate start in $B$.
This leads to a guaranteed $O(w)$ with {\sc FindStart}().
In practice, the two algorithms have almost identical runtime.
We use Algorithm~\ref{algo:LC1} in longdust as it is closer to the exact algorithm.

Function {\sc FindStart}() finds the longest good LC interval ending at one position.
We apply the function to every position in the genome to find all good LC intervals.
We can skip $j$ if $[j-k+1,j]$ is unique in $[j-w+1,j]$ because the forward pass would not reach $j$ in this case.
We also introduce a heuristic to extend a good LC interval $[j-w,j-1]$ to $[j-w+1,j]$ without calling {\sc FindStart}().

Our definition may classify a short non-repetitive interval to LC if it is surrounded by highly repetitive LC intervals.
We additionally use an X-drop heuristic~\citep{Altschul:1997vn} to alleviate this issue.
On the T2T-CHM13 genome, the total length of LC intervals is 0.5\% shorter with X-drop.
The heuristic has a minor effect.

The overall longdust algorithm is inexact and may result in slightly different LC regions on opposite strands.
We run the algorithm on both the forward and the reverse strand of the input sequences and merge the resulting intervals.
The default longdust output is strand symmetric.

\subsection{The SDUST scoring}

SDUST~\citep{Morgulis:2006aa} uses the following complexity scoring function:
$$
S_S(\vec{c}_x)=\frac{1}{\ell(x)}\sum_t\frac{c_x(t)(c_x(t)-1)}{2}-T
$$
This function grows linearly with $\ell(x)$ for $\ell(x)\ge4^k$,
while our scoring function grows more slowly in the logarithm scale.
The SDUST function is more likely to classify longer sequences as LC.

Furthermore, SDUST looks for perfect LC intervals rather than good LC intervals like longdust.
It cannot test whether an interval is perfect in linear time.
Instead, SDUST maintains the complete list of perfect intervals in window $[j-w+1,j]$
and tests a new candidate interval against the list.
The {\sc FindStart}() equivalent of SDUST is $O(w^3)$ in time, impractical for long windows.
SDUST hardcodes $k=3$ and uses $w=64$ by default for acceptable performance.

\subsection{Scoring with Shannon entropy}

Let $p_x(t)=c_x(t)/\ell(x)$.
The Shannon entropy of string $x$ is
$$
H(\vec{c}_x)\triangleq-\sum_tp_x(t)\log p_x(t)=\log\ell(x)-\frac{1}{\ell(x)}\sum_t c_x(t)\log c_x(t)
$$
When $\ell(x)\le4^k$, $H(\vec{c}_x)$ reaches the maximum value of $\log\ell(x)$ at $p_x(t)=1/\ell(x)$.
$H(\vec{c}_x)$ also grows with $\ell(x)$.
For $\ell(x)\le4^k$, define
$$
S_E(\vec{c}_x)\triangleq\log\ell(x)-H(\vec{c}_x)-T=\frac{1}{\ell(x)}\sum_t c_x(t)\log c_x(t)-T
$$
We adapted longdust for $S_E$ and found using $S_E$ is more than twice as slow.
We suspected some longdust heuristics did not work well with $S_E$.
Not backed up by a statistical model, it is not apparent how to integrate GC adjustment, either.

\section{Results}

\subsection{Low-complexity regions in T2T-CHM13}

\begin{table}[tb]
\caption{Command lines and resource usage for T2T-CHM13\label{tab:cmd}}
\begin{tabular*}{\columnwidth}{@{\extracolsep\fill}lrrl@{\extracolsep\fill}}
\toprule
Tool & $t_{\rm CPU}$ (h) & Mem (G) & Command line \\
\midrule
longdust & 0.94   & 0.47 & (default) \\
SDUST    & 0.07   & 0.23 & {\tt -t30} \\
pytrf    & 2.64   & 0.70 & {\tt -M500} \\
TRF      & 12.83  & 7.49 & {\tt 2 7 7 80 10 50 500 -l12} \\
TANTAN   & 0.56   & 1.28 & {\tt -w500} \\
ULTRA    & 146.07 & 33.31& {\tt -p 500 -t 16} \\
\botrule
\end{tabular*}
\begin{tablenotes}\setlength\itemsep{0.0em}
Performance measured on a Linux server equipped with Intel Xeon Gold 6130 CPU and 512GB memory.
\end{tablenotes}
\end{table}

We applied longdust, SDUST v0.1~\citep{Morgulis:2006aa},
pytrf v1.4.2~\citep{Du:2025aa},
TRF v4.10~\citep{Benson:1999aa},
TANTAN v51~\citep{Frith:2011aa},
and ULTRA v1.20~\citep{Olson:2024aa}
to the T2T-CHM13 human genome~\citep{Nurk:2022up}.
The maximum period was set to 500 (Table~\ref{tab:cmd}).
Notably, TRF would not finish in days with the default option {\tt -l}.
With {\tt -l30}, TRF could run in 11 hours using 17.82GB memory at the peak.

\begin{figure}[tb]
\includegraphics[width=\columnwidth]{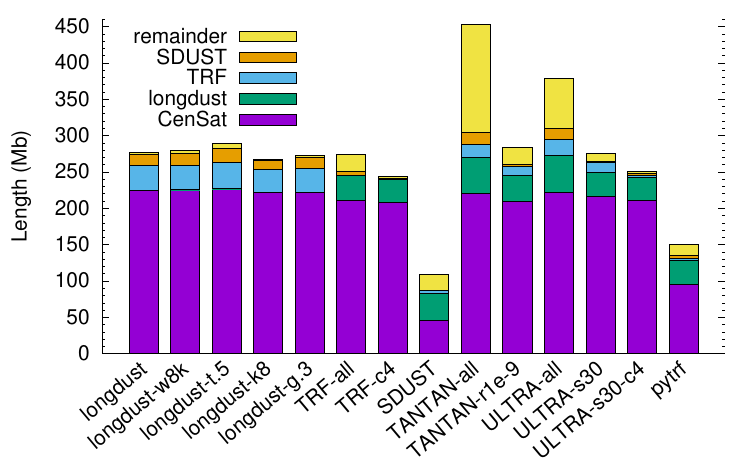}
\caption{Lengths of low-complexity regions in T2T-CHM13.
Low-complexity (LC) regions identified by each tool are first intersected with centromeric satellite annotation.
The remainder is then intersected with longdust, TRF and SDUST in order.
There are no overlaps between stacks.
The total height is the length of LC regions found by each tool.
Alternative settings --
``longdust-w8k'': window size $w=8000$;
``longdust-t.5'': score threshold $T=0.5$;
``longdust-k8'': using 8-mers;
``longdust-g.3'': genome-wide GC set to 30\%;
``TRF-c4'': requiring $\ge4$ copies of the repeat unit;
``TANTAN-r1e-9'': run with ``-r 1e-9'';
``ULTRA-s30'': requiring score $\ge30$ in the output.
}\label{fig:len}
\end{figure}

Due to the lack of ground truth, we could not calculate the accuracy of longdust.
We instead checked how often longdust reports regions unique to itself
and how often it misses regions reported by other tools, especially by two or more of them.

Longdust finds 277.1Mb of LC regions with 224.3Mb overlapping with centromeric satellite annotated by the telomere-to-telomere (T2T) consortium (Fig.~\ref{fig:len}).
Of the remaining 52.7Mb, 34.1Mb overlaps with TRF; 15.4Mb of the remainder (18.6Mb) is found by SDUST.
Only 3.2Mb is left, suggesting most longdust LC regions fall in centromeres or are found by TRF or SDUST.
Longdust uses three parameters: $k$-mer size, window size $w$ and score threshold $T$.
Varying these parameters do not greatly alter the results.
The GC adjustment also has minor effect even when we intentionally set the genome-wide GC below the actual value of 41\%.

TRF, the most popular tandem repeat finder, finds 274.5Mb of tandem repeats,
244.0Mb of which have $\ge4$ copies of repeat units.
97.9\% of the 244.0Mb are identified by longdust.
TRF additionally reports tandem repeats with $<4$ repeat units.
Only 14.8\% of them overlap with longdust results.
Longdust misses repeats with low copy numbers.
In fact, under the default threshold $T=0.6$,
the minimum numbers of exact copies longdust can find is approximately:
$$
3+\frac{k-1}{r}+\frac{3T-\log2-\log3}{\log4-T}\approx3.01+\frac{k-1}{r}
$$
This assumes the repeat unit length $r>k$, $f(\cdot)=0$ and all $k$-mers are unique within the repeat unit.

ULTRA outputs 68.2Mb of regions not reported by longdust, TRF or SDUST.
Nevertheless, if we raise its score threshold to 30,
the total region length is similar to that of TRF
with most of repeats of $\ge$4 copies captured by longdust.
Under the default setting, TANTAN finds the largest LC regions missed by longdust, TRF or SDUST.
Decreasing the repeat starting threshold from 0.005 to $10^{-9}$ greatly reduces its unique regions from 148.5Mb to 24.5Mb.
We suspect the remaining regions unique to TANTAN may have low copy number but we cannot control its output.
TANTAN can optionally generate repeat units but this mode is six times as slow and reports only 27\% of regions in the default mode.

As to other tools, pytrf cannot effectively identify alpha satellite with $\sim$170bp repeat units,
even though it was set to find tandem repeats with unit up to 500bp.
With a 64bp window size, SDUST naturally misses tandem repeats with long units, including all alpha satellite.

\subsection{Behaviors in long satellite repeats}

Although longdust, TRF, TANTAN and ULTRA report similar amount of human satellites,
the LC block sizes vary greatly.
We will take the centromeric satellite of chromosome 11 for an example.
Around this centromere, longdust reports a continuous block of 3.5Mb from 50,999,020 to 54,488,059 on T2T-CHM13.
TRF finds a nearly identical region, with the ending position shifted by only 3bp.
In contrast, although ULTRA finds 99.86\% of this region,
it breaks them into many smaller blocks.
The longest block is 120kb in size.
TANTAN under the default setting misses 2.3\% of this long alpha array and its output is more fragmented.
The longest block is only 1587bp.
With ``-r 1e-9'', TANTAN misses 7.6\% of this satellite.
Across all satellite regions annotated on T2T-CHM13,
the N50 block size is 2.9Mb, 2.0Mb, 10.2kb and 340bp for longdust, TRF, ULTRA and TANTAN, respectively.
Longdust reports the longest blocks probably because it merges satellites of different types.

One application of LC masking is to speed up cross-species whole-genome alignment.
90.8\% of satellites found by longdust lie in blocks longer than 10kb
and conversely, 99.2\% of blocks longer than 10kb fall in annotated satellites.
By setting a block length threshold, we can mask the bulk of satellite without affecting the rest of the genome.
A similar strategy works for TRF that also reports long blocks.
We would need to stitch shorter blocks reported ULTRA and TANTAN to use this strategy,
which would add another layer of complexity.

\subsection{Checking strand symmetry}

\begin{figure}[tb]
\includegraphics[width=\columnwidth]{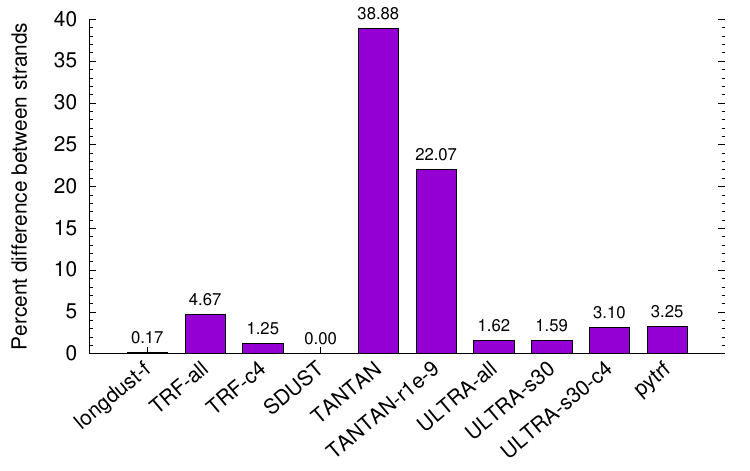}
\caption{Strand asymmetry.
Each tool is run on the original T2T-CHM13 and the reverse complement of the genome.
The LC intervals outputted from the two runs are compared.
The Y axis shows ${\rm 1-intersection/union}$ for each tool.
Longdust by default applies the algorithm to both strands and outputs the union;
``longdust-f'' refers to the mode that processes one strand only.
}\label{fig:strand}
\end{figure}

SDUST is strand symmetric in that running SDUST on either strand of the input sequences gives identical results.
While longdust defines string complexity with strand symmetry,
it uses heuristics in implementation and may report different LC intervals if it is run on the reverse complement of the input sequences.
To achieve strand symmetry, longdust applies the same algorithm to both strands and output the union intervals.
Without the union, 0.17\% of LC regions are specific to one strand (Fig.~\ref{fig:strand}).

We also checked the strand differences of other tools.
For most of them, a few percent of LC regions are strand specific.
However, a large fraction of LC regions found by TANTAN was only reported on one strand.
The difference is driven by regions outside centromeric satellites, where the percent difference reaches 60.0\%.

\subsection{Low-complexity regions in a gorilla genome}

The near T2T gorilla genome (AC:GCF\_029281585.2; \citealp{Yoo:2025aa}) is 3546Mb in size, 428Mb larger than the human T2T-CHM13 genome.
We ran longdust on the gorilla genome for 1.4 hours and found 656.8Mb of LC regions.
That is 379.7Mb larger than the LC regions in T2T-CHM13.
The genome size difference is primarily driven by LC regions.

To further confirm this observation,
we extracted 298.8Mb of regions in the gorilla genome that are $\ge10$kb in length without any 51-mer exact matches to 472 human genomes~\citep{Li:2024ac}.
99.7\% of them are marked as LC regions by longdust and none of them are alpha satellite.
95.8\% of these gorilla-specific regions are distributed within 15Mb from telomeres,
broadly in line with \citet{Yoo:2025aa}.

TANTAN with ``{\tt -w500 -r1e-9}'' reported 672.2Mb of LC regions, 629.1Mb of which overlapped with longdust.
TANTAN still tends to report fragmented regions with an N50 block length of 3.7kb.
We also ran TRF on the gorilla genome.
It did not finish in a week with option ``{\tt -l20}''.

\section{Discussions}


\citet{Du:2025aa} observed that each tandem repeat finding tool reported a large fraction of regions unique to itself.
We think this is driven by weak tandem repeats that are composed of a few diverged repeat units.
If we focus on highly repetitive LC regions, longdust, TRF, TANTAN and ULTRA agree reasonably well with each other.
It is difficult to tell which tool is more accurate in terms of genome-wide coverage.
We do not intend to brand longdust as the best tool for LC finding;
we emphasize more on a new probabilistic model for defining string complexity,
which advances the SDUST algorithm and achieves a descent balance between performance, coverage, block contiguity and strand symmetry.

From the theoretical point of view, longdust uses an approximate algorithm.
It tests LC intervals ending at each position with Algorithm~\ref{algo:LC1},
but has not sufficiently exploited dependencies between positions.
It will be interesting to see whether there is an exact $O(wL)$ algorithm under the longdust formulation
or a meaningful alternative formulation that leads fast implementations.

In comparison to tandem repeat finding tools such as TRF and ULTRA,
longdust is unable to report the repeat units in case of tandem repeats.
This will limit the use cases of longdust.
Another practical limitation of longdust is the restricted window size.
The genome of Woodhouse's scrub jays, for example, contains satellite with a 18kb repeat unit~\citep{Edwards:2025aa}.
This would be missed by longdust under the default setting.
Increasing the window size would make longdust considerably slower.
This is partly due to the $O(wL)$ time complexity and partly due to the speedup strategies in longdust that are more effective given $\ell(x)\ll 4^k$.
It would be ideal to have an algorithm that remains efficient given large windows or, better, does not require specified window sizes.

\section*{Acknowledgments}

We are grateful to Qian Qin for evaluating the effect of low-complexity regions in structural variant calling,
and to Maximilian Haeussler for explaining the role of masking low-complexity regions in cross-species alignment.
We thank the reviewers for suggesting cleaner derivation.

\section*{Author contributions}

H.L. conceived the project, implemented the method, analyzed the data and drafted the manuscript.
B.L. prototyped the algorithm.

\section*{Conflict of interest}

None declared.

\section*{Funding}

This work is supported by US National Institute of Health grant R01HG010040, R01HG014175, U01HG013748, U41HG010972, and U24CA294203 (to H.L.).

\section*{Data availability}

\url{https://github.com/lh3/longdust}

\bibliographystyle{apalike}
{\sffamily\small
\bibliography{longdust}}

\end{document}